\documentclass[10pt,twocolumn,floatfix,superscriptaddress,amsmath,aps,pre,eqsecnum]{revtex4-2}
\usepackage[utf8]{inputenc}
\usepackage[T1]{fontenc}
\usepackage{lmodern}
\usepackage{amsmath,amssymb,mathtools}
\usepackage{graphicx,booktabs,microtype}
\usepackage{xcolor}
\usepackage{placeins}
\usepackage[unicode,hidelinks]{hyperref}
\hypersetup{
 pdftitle={Geometric properties of Wilsonian flows of periodic potentials in two dimensions},
 pdfauthor={Daniel G. Barci and Nathan O. Silvano},
 pdfsubject={Geometric evolution of periodic potentials in Wilsonian renormalization group}
}
\newcommand{\dd}{\mathrm{d}}
\newcommand{\pd}{\partial}
\newcommand{\av}[1]{\langle #1\rangle_{>}}
\newcommand{\Sp}{\mathcal S}

\usepackage[normalem]{ulem}
\begin{document}
\title{\bfseries Geometric properties of Wilsonian flows of periodic potentials in two dimensions}
\author{Daniel G. Barci}
\affiliation{Departamento de F{\'\i}sica Te\'orica, Universidade do Estado do Rio de Janeiro,
Rua S\~ao Francisco Xavier 524, 20550-013, Rio de Janeiro, RJ, Brazil.}
\author{Nathan O. Silvano}
\affiliation{Instituto de F\'isica Interdisciplinar y Sistemas Complejos (IFISC), CSIC-UIB,
Campus Universitat de les Illes Balears, E-07122 Palma de Mallorca, Spain.}
\date{\today}

\begin{abstract}
Periodic interactions can possess multiple competing minima and barriers, whose evolution under coarse-graining offers a geometric perspective complementary to the scaling of Fourier harmonics. We study this evolution in two dimensions by deriving a functional Wilsonian renormalization-group flow for a general smooth periodic potential and its running sine-Gordon frequency within a constant-kinetic projection. Reaction--diffusion and Burgers representations reveal two key properties of the flow: discrete translation symmetries are preserved, and the number of extrema can only remain constant or decrease. Consequently, the nonlinear generation of Fourier harmonics may accompany a progressive simplification of the potential landscape. We illustrate the role of symmetry with two complementary examples. A weak breaking of the $\pi$-translation symmetry allows a reduction from four extrema to two, whereas a non-sinusoidal interaction retaining this symmetry evolves from eight extrema to the four protected by its periodicity. These results highlight how symmetry and coarse-graining jointly reorganize competing minima and barriers in periodic interactions.
\end{abstract}

\maketitle

\section{Introduction}
Periodic interactions arise in many physical settings, including crystalline interfaces, commensurate structures, substrate pinning, and bosonized descriptions of one-dimensional quantum systems. The sine-Gordon model provides their simplest representation, while more general periodic potentials can exhibit several local minima within a fundamental period, inequivalent barriers, and distinct local curvatures. Such structures occur, for example, in models of surface preroughening, multifrequency sine-Gordon theories, and deformable substrate potentials~\cite{Prestipino1999,Fabrizio2000,Peyrard1982}. Following how these features evolve as short-wavelength fluctuations are integrated out provides a natural way to characterize the coarse-graining of general periodic interactions.

The renormalization of periodic interactions has a long history in connection with the Kosterlitz--Thouless (KT) transition and its descriptions in the two-dimensional XY model, Coulomb gas, and sine-Gordon theory~\cite{KosterlitzThouless1973,Kosterlitz1974,Jose1977,Amit1980}. In this context, the relevance of periodic perturbations and the stability of the Gaussian fixed line are conveniently characterized by the scaling dimensions of Fourier harmonics~\cite{Jose1977,Amit1980,Malard2013}. Momentum-shell calculations and subsequent differential and functional RG approaches have also addressed more general periodic interactions~\cite{Ohta1978,Knops1980,Nandori2001,Nandori2002}. Further functional-RG and critical-flow studies have enriched our understanding of sine-Gordon renormalization~\cite{Nagy2009,PelissettoVicari2013,DavietDupuis2019}, including recent comparisons between perturbative and nonperturbative RG and alternative Wilson--Polchinski formulations~\cite{Hariharakrishnan2024,LiVicedo2026}. This broader framework provides the starting point for studying how the shape of a periodic interaction evolves along an RG trajectory.

Our focus is the \emph{evolution of the shape of a periodic potential} as short-wavelength fluctuations are successively integrated out. At each coarse-graining scale, the running potential describes an effective interaction between slowly varying field configurations. The number and arrangement of its maxima and minima, the barriers separating them, and the discrete symmetries relating equivalent configurations offer a geometric characterization of this evolution. Following these properties across scales reveals how a microscopic potential with several competing extrema is reorganized into an effective long-distance interaction.

Fourier harmonics provide a natural description of the scaling of periodic perturbations. At the same time, the global shape of the potential contains information about the number and arrangement of its extrema. Nonlinear RG equations couple different harmonics and generate additional Fourier components along the flow~\cite{Yanagisawa2021}. An important possibility is that this increasing harmonic content accompanies a simpler geometric landscape, with fewer extrema. Keeping the complete field dependence of the running potential allows us to examine these two aspects of the evolution together.

In this work, we study the Wilsonian RG flow of a general smooth periodic potential in two dimensions. Using a smooth momentum cutoff, we integrate out short-wavelength modes and derive the flow equations through a second-order cumulant expansion. A local derivative projection then gives a closed functional equation for the potential, together with the flow of the sine-Gordon frequency. We show that our resulting equations agree with the field independent kinetic projection of the derivative-expanded Polchinski RG equation~\cite{Nandori2002}, once the regulator and field-normalization conventions are matched. This agreement provides a link between the explicit mode-elimination procedure and the differential RG formulation.
	
	We use the resulting flow equation to investigate global geometric changes of the potential at different scales. We find that potentials with discrete translation symmetries retain those symmetries throughout the RG flow. We also show, using a scalar parabolic Burgers equation, that the number of extrema cannot increase: it either stays the same or decreases as the flow proceeds. These two results imply that the translation symmetry sets a lower bound on the number of extrema that can survive along the flow. Additional extrema may disappear through mergers, but the symmetry-related structure remains. We illustrate this behavior with two examples: a weakly symmetry-broken potential whose number of extrema decreases from four to two, and a non-sinusoidal potential that evolves from eight to four extrema while preserving its $\pi$-translation symmetry. Our results show how the global shape of a periodic potential evolves under RG transformations, providing a geometric complement to the usual analysis in terms of Fourier harmonics.

\section{Smooth-cutoff Wilsonian flow}
We consider the Euclidean Hamiltonian in two dimensions,
\begin{equation}
 H[\phi]=\int\dd^2x\left[\frac12(\nabla\phi)^2
 -\frac12 V(\beta\phi)\right],
 \label{eq:ham-dimensional}
\end{equation}
where $V(\theta)$ is a smooth real periodic potential of the angular variable $\theta=\beta\phi$, satisfying $V(\theta+2\pi)=V(\theta)$. In two dimensions, $V$ has dimensions of squared momentum. For the renormalization-group analysis, it is convenient to introduce the dimensionless periodic potential $u(\theta)$ through $V(\theta)=\Lambda^2 u(\theta)$, where $\Lambda$ is the reference momentum cutoff. The Hamiltonian then takes the form
\begin{equation}
 H[\phi]=\int\dd^2x\left[\frac12(\nabla\phi)^2
 -\frac{\Lambda^2}{2}\,u(\beta\phi)\right],
 \label{eq:ham}
\end{equation}
with
\begin{equation}
 u(\theta+2\pi)=u(\theta).
 \label{eq:periodicity}
\end{equation}
We do not restrict the Fourier content of $u$. Its constant part contributes only to the field-independent free energy and does not affect the geometric properties considered below. The sine-Gordon frequency $\beta$ fixes the period $2\pi/\beta$ in the canonically normalized field $\phi$, whose kinetic coefficient is set to unity. In the angular variable $\theta$, the Gaussian term is $(2\beta^2)^{-1}(\nabla\theta)^2$, so the Gaussian stiffness is $\beta^{-2}$ in our normalization. Allowing $\beta$ to run is therefore equivalent to renormalizing this stiffness while keeping the angular period of $u$ fixed.

We employ a Gaussian smooth-cutoff covariance,
\begin{equation}
 C_\Lambda(q)=\frac{e^{-q^2/\Lambda^2}}{q^2}.
 \label{eq:regcov}
\end{equation}
For an infinitesimal RG step $\Lambda\mapsto \Lambda e^{-\dd\ell}$, the covariance of the integrated modes is defined by the difference
\begin{equation}
 G_{>}(q)=C_\Lambda(q)-C_{\Lambda e^{-\dd\ell}}(q).
 \label{eq:shellcov}
\end{equation}
Expanding this difference to first order in $\dd\ell$, we obtain
\begin{equation}
 G_{>}(q)=\frac{2\dd\ell}{\Lambda^2}
 \exp\!\left(-\frac{q^2}{\Lambda^2}\right)+O(\dd\ell^2).
 \label{eq:fastcov}
\end{equation}
Although $C_\Lambda(q)$ is singular at $q=0$, the difference~\eqref{eq:shellcov} is regular there. In particular, its Taylor expansion in the \emph{external} momentum $q$ is well defined,
\begin{equation}
 G_{>}(q)=\frac{2\dd\ell}{\Lambda^2}
 \left[1-\frac{q^2}{\Lambda^2}+O\!\left(\frac{q^4}{\Lambda^4}\right)\right]
 +O(\dd\ell^2).
 \label{eq:smallq}
\end{equation}
The variance of the integrated modes is
\begin{equation}
 G_{>}^{(x)}(0)
 \equiv\int\frac{\dd^2q}{(2\pi)^2}G_{>}(q)
 =\frac{\dd\ell}{2\pi}+O(\dd\ell^2).
 \label{eq:Gzero}
\end{equation}
Here $G_{>}^{(x)}$ denotes the position-space Fourier transform of $G_{>}(q)$, so $G_{>}^{(x)}(0)$ is the coincident-point variance.

We split the field into slow and integrated components,
\begin{equation}
 \phi=\varphi+\chi,
\end{equation}
with the interaction term
\begin{equation}
 H_I[\varphi+\chi]
 =-\frac{\Lambda^2}{2}\int\dd^2x\,u(\beta\varphi+\beta\chi).
\end{equation}
The effective Hamiltonian is obtained from the cumulant expansion,
\begin{equation}
 H_{\mathrm{eff}}=H_0[\varphi]+\av{H_I}
 -\frac12\bigl(\av{H_I^2}-\av{H_I}^2\bigr)+\cdots.
 \label{eq:cumulants}
\end{equation}
With $\theta(x)=\beta\varphi(x)$, the first cumulant is a Gaussian convolution of $u$ in its argument,
\begin{equation}
 \av{u(\theta+\beta\chi)}
 =u(\theta)+\frac{\beta^2}{2}\,G_{>}^{(x)}(0)\,u''(\theta)
 +O(\dd\ell^2),
 \label{eq:firstc}
\end{equation}
where the prime ($'$) indicates derivatives with respect to the field $\theta$.

To second order in the periodic interaction, the connected cumulant is
\begin{equation}
 \begin{aligned}
  \Delta H^{(2)}&=-\frac12\bigl(\av{H_I^2}-\av{H_I}^2\bigr)
 \\
 &=-\frac{\Lambda^4\beta^2}{8}
 \int\dd^2x\,\dd^2y\,G_{>}(x-y)\\
 &\quad\times u'(\theta(x))u'(\theta(y))
 +O(u^3,u^2\dd\ell^2).
 \end{aligned}
 \label{eq:second}
\end{equation}
In order to make a gradient expansion this non-lolcal  expression, we express this convolution in momentum space. We introduce the Fourier transform of the composite slow-field function,
\begin{equation}
 F(q)=\int\dd^2x\,e^{-iq\cdot x}u'(\theta(x)).
 \label{eq:F}
\end{equation}
The second cumulant then takes the compact form
\begin{equation}
 \begin{aligned}
 \Delta H^{(2)}={}&-\frac{\Lambda^4\beta^2}{8}
 \int\frac{\dd^2q}{(2\pi)^2}\,G_{>}(q)\\
 &\quad\times F(q)F(-q)+O(u^3,u^2\dd\ell^2).
 \end{aligned}
 \label{eq:secondFourier}
\end{equation}
Substituting the small-momentum expansion~\eqref{eq:smallq}, we obtain the leading local and two-derivative terms,
\begin{align}
 \Delta H^{(2)}={}&-\frac{\Lambda^2\beta^2}{4}\,\dd\ell
 \int\frac{\dd^2q}{(2\pi)^2}F(q)F(-q)\notag\\
 &+\frac{\beta^2}{4}\,\dd\ell
 \int\frac{\dd^2q}{(2\pi)^2}q^2F(q)F(-q)\notag\\
 &+O(\partial_x^4,u^3,\dd\ell^2).
 \label{eq:momentumExpansion}
\end{align}
Parseval's identity identifies the first momentum integral with the local square of the first derivative of the potential,
\begin{equation}
 \int\frac{\dd^2q}{(2\pi)^2}F(q)F(-q)
 =\int\dd^2x\,[u'(\theta)]^2.
 \label{eq:parseval0}
\end{equation}
Similarly, the quadratic moment corresponds to the squared spatial gradient of the same function,
\begin{equation}
 \int\frac{\dd^2q}{(2\pi)^2}q^2F(q)F(-q)
 =\int\dd^2x\,\bigl|\nabla u'(\theta)\bigr|^2.
 \label{eq:parseval2}
\end{equation}
Since $\theta=\beta\varphi$, the chain rule yields
\begin{equation}
 \bigl|\nabla u'(\theta)\bigr|^2
 =\beta^2[u''(\theta)]^2(\nabla\varphi)^2.
 \label{eq:chain}
\end{equation}
The local derivative expansion of the second cumulant is therefore
\begin{align}
 \Delta H^{(2)}={}&-\frac{\Lambda^2\beta^2}{4}\,\dd\ell
 \int\dd^2x\,[u'(\theta)]^2 
  \label{eq:localSecond}  \\
 &+\frac{\beta^4}{4}\,\dd\ell
 \int\dd^2x\,[u''(\theta)]^2(\nabla\varphi)^2
 +O(\partial_x^4,u^3,\dd\ell^2).
\nonumber 
\end{align}

The second line of Eq.~\eqref{eq:localSecond} generates a \emph{field-dependent} kinetic correction. To obtain a closed flow for a periodic potential and its running sine-Gordon frequency, we retain its field-independent component and discard the remaining kinetic harmonics and higher-gradient operators. The corresponding correction to the Gaussian kinetic coefficient is
\begin{equation}
 \delta Z=\frac{\beta^4\,\dd\ell}{2}
 \frac{1}{2\pi}\int_0^{2\pi}\dd\theta\,[u''(\theta)]^2.
 \label{eq:deltaZ}
\end{equation}
We restore the canonical kinetic normalization by rescaling the slow field, which changes the parameter $\beta$ according to
\begin{equation}
 \beta\longmapsto\frac{\beta}{\sqrt{1+\delta Z}}.
\end{equation}
The spatial rescaling that restores the reference cutoff then gives the flow of the complete periodic potential,
\begin{equation}
 \pd_\ell u(\theta,\ell)=2u+D(\ell)u''
 +\frac{\beta^2}{2}(u')^2,
 \label{eq:u}
\end{equation}
where the diffusion coefficient is
\begin{equation}
 D(\ell)=\frac{\beta^2(\ell)}{4\pi}.
 \label{eq:D}
\end{equation}
The accompanying flow of the sine-Gordon frequency is
\begin{equation}
 \frac{\dd\beta}{\dd\ell}=-\frac{\beta^5}{8\pi}
 \int_0^{2\pi}\dd\theta\,[u''(\theta,\ell)]^2.
 \label{eq:beta}
\end{equation}
 These equations hold through second order in the interaction within the specified derivative projection. The coefficient of the linear $u''$ term is fixed by the Gaussian shell variance and is independent of the detailed profile of a regular, normalized smooth cutoff. Indeed, for $C_\Lambda(q)=K(q^2/\Lambda^2)/q^2$ with $K(0)=1$ and $K(\infty)=0$, the infinitesimal variance remains $G_>^{(x)}(0)=\dd\ell/(2\pi)$ in two dimensions. By contrast, the nonlinear coefficients and the flow of the projected kinetic term can depend on the cutoff profile and the projection scheme~\cite{DePolsiWschebor2022,Hariharakrishnan2024}.

The coupled functional equations~\eqref{eq:u} and~\eqref{eq:beta} are compatible with the derivative-expanded smooth-cutoff Polchinski RG for periodic scalar theories~\cite{Nandori2002}. This agreement provides an independent check on the coefficients obtained by the direct Wilsonian cumulant method. The correspondence requires projecting the field-dependent wave-function renormalization onto its constant Fourier component and restoring the canonical kinetic normalization after each blocking step. In Appendix~\ref{app:polchinski}, we give the explicit mapping, explain why nonconstant kinetic harmonics are linearly irrelevant near the Gaussian fixed line, and examine the limitations of the projection.

The Gaussian scaling dimensions and their associated KT thresholds are most directly recovered from the reaction--diffusion representation developed below.

\section{Functional evolution: reaction--diffusion and Burgers representations}

The closed equation~\eqref{eq:u} evolves a complete periodic function.  Its nonlinear derivative term obscures two useful features: the relation to diffusion in the periodic field coordinate and the restrictions on the evolution of local extrema. Two changes of variables bring these structures into focus. We first examine a positive transformation of the potential and its reaction--diffusion dynamics, and subsequently turn to the derivative of the potential, whose zeros carry the geometric information.

\subsection{Reaction--diffusion equation for the transformed potential}

We introduce the positive function
\begin{equation}
 W(\theta,\ell)=\exp[2\pi u(\theta,\ell)]>0.
 \label{eq:Wdef}
\end{equation}
This exponential substitution is a Cole--Hopf transformation: it eliminates the quadratic-gradient nonlinearity of the potential flow, while leaving a nonlinear reaction term.
The coefficient $2\pi$ is selected so that the quadratic term in the derivative of $u$ cancels against the derivative term generated by $W''$. More explicitly, since
\begin{equation}
 \frac{W''}{W}=2\pi u''+4\pi^2(u')^2,
 \label{eq:Widentity}
\end{equation}
Eq.~\eqref{eq:u} becomes
\begin{equation}
 \partial_\ell W=D(\ell)W''+2W\ln W,
 \label{eq:W}
\end{equation}
with $D(\ell)=\beta^2(\ell)/(4\pi)>0$. This is a reaction--diffusion equation on the angular circle~\cite{Smoller1994}.  Of course, it is not a diffusion process in physical space or physical time. Its evolution variable is the logarithmic RG scale $\ell$, whereas diffusion acts along $\theta=\beta\phi$, the argument of the periodic interaction. The running sine-Gordon frequency in Eq.~\eqref{eq:beta} makes the diffusion coefficient scale dependent, but not dependent on $\theta$ within our projection.

The diffusion term smooths variations of $W$ in the angular field coordinate. For the chosen additive normalization of $u$, the local reaction $2W\ln W$ is positive above $W=1$ and negative below it. This threshold depends on  a normalization of the physical boundary: a field-independent shift $u\mapsto u+c(\ell)$ rescales $W$ and changes its evolution by an additional term linear in $W$ with a coefficient uniform in $\theta$. However, the periodicity sectors and locations of extrema, which are the geometric observables of interest, are unaffected. The choice $u=0$, or $W=1$, is a convenient uniform reference configuration. Positivity follows from the exponential transformation for any finite smooth solution. The local reaction tendency should not be confused with the evolution of individual Fourier amplitudes, which are also affected by diffusion.

It is interesting to study the competition between diffusion and reaction  near this uniform configuration. Writing $W=1+\delta W$ with $|\delta W|\ll1$ gives
\begin{equation}
 \partial_\ell\delta W=D(\ell)\delta W''+2\delta W
 +O((\delta W)^2).
 \label{eq:Wlinear}
\end{equation}
A Fourier perturbation $\delta W\propto e^{in\theta}$ therefore has the instantaneous Gaussian scaling exponent
\begin{equation}
 \lambda_n=2-Dn^2=2-\frac{n^2\beta^2}{4\pi}.
 \label{eq:lambdaW}
\end{equation}
Its marginality condition reproduces the familiar hierarchy
\begin{equation}
 \beta_{c,n}^{2}=\frac{8\pi}{n^2}.
 \label{eq:Wthresholds}
\end{equation}
Thus the reaction--diffusion equation contains the standard KT linear thresholds, but it also describes the nonlinear evolution of the full potential far beyond the single-harmonic ansatz.

An immediate consequence is the preservation of discrete translation symmetries. For each positive integer $p$, the space of functions with period $2\pi/p$ can be written as
\begin{equation}
 \Sp_p=\{f\in C^2(\mathbb S^1):
 f(\theta+2\pi/p)=f(\theta)\}.
 \label{eq:Sp}
\end{equation}
If an initial potential belongs to $\Sp_p$, its transformed profile $W$ has the same period. Both $W(\theta,\ell)$ and $W(\theta+2\pi/p,\ell)$ satisfy Eq.~\eqref{eq:W} with the same initial condition; the accompanying equation for $\beta$ is also unchanged by the translation. Uniqueness of the smooth initial-value problem consequently ensures that the discrete translation symmetry persists at every finite scale at which the solution exists.

These sectors are nested, for instance $\Sp_1\supset\Sp_2\supset\Sp_4$. They are not independent Fourier sectors for a general potential: nonlinearities generate new harmonics compatible with the smallest period already present. The scaling relation
\begin{equation}
 u_p(\theta,\ell)=\widetilde u(p\theta,\ell),
 \label{eq:covering-u}
\end{equation}
accompanied by
\begin{equation}
 \beta_p(\ell)=\frac{\widetilde\beta(\ell)}{p},
 \label{eq:covering-beta}
\end{equation}
relates solutions in different periodicity sectors. It also makes the $1/p^2$ dependence of the Gaussian thresholds natural. These statements express covariance under discrete field translations within the projected flow; they do not require stationary RG trajectories.

The stationary auxiliary reaction--diffusion problem also admits a mechanical interpretation. With the above normalization of $W$ and with $\beta$ held fixed and setting the left-hand side of Eq.~\eqref{eq:W} to zero gives an auxiliary stationary problem,
\begin{equation}
 D W''=-\frac{\dd\mathcal U}{\dd W},
 \label{eq:aux-oscillator}
\end{equation}
where
\begin{equation}
 \mathcal U(W)=\frac12W^2\bigl(\ln W^2-1\bigr),\qquad W>0.
 \label{eq:aux-potential}
\end{equation}
If $\theta$ is viewed as the time of a fictitious particle of mass $D$, Eq.~\eqref{eq:aux-oscillator} is Newton's equation in a logarithmic anharmonic potential. It admits the first integral
\begin{equation}
 \mathcal E=\frac{D}{2}(W')^2+\mathcal U(W),
 \label{eq:oscillator-energy}
\end{equation}
which reduces the stationary auxiliary problem to a quadrature. Small oscillations about the positive minimum $W=1$ have frequency
\begin{equation}
 \omega_0^2=\frac{2}{D}=\frac{8\pi}{\beta^2}.
 \label{eq:aux-frequency}
\end{equation}
The condition of $p$ small oscillations within the angular period, $\omega_0=p$, identifies the same marginality condition as Eq.~\eqref{eq:Wthresholds}. The oscillator interpretation thus provides an intuitive reading of the integer hierarchy derived from Eq.~\eqref{eq:Wthresholds}. It concerns stationary solutions of the \emph{fixed-$\beta$ auxiliary equation}, not nontrivial fixed points of the coupled RG system. Indeed, for $\beta>0$, Eq.~\eqref{eq:beta} has a vanishing right-hand side only if $u''=0$ and periodicity then forces $u$ to be constant.

The auxiliary oscillator concerns this particular normalization of the reaction term. A different choice for the field-independent part of the potential modifies the stationary auxiliary equation but not the symmetry or extremum-counting results that follow.

\subsection{Burgers equation and the evolution of extrema}

The geometry of maxima and minima of the potential  is more directly expressed through the first derivative of the potential,
\begin{equation}
 v(\theta,\ell)=u'(\theta,\ell).
 \label{eq:vdef}
\end{equation}
Because $u$ is periodic, $v$ is periodic and has zero angular average. Differentiating Eq.~\eqref{eq:u} gives
\begin{equation}
\partial_\ell v=D(\ell)v''+\beta^2(\ell)v v'+2v.
 \label{eq:v}
\end{equation}
The term proportional to $v v'$ has the familiar nonlinear advective form of the viscous Burgers equation~\cite{Whitham1999}; the additional linear term acts as a local source, while the positive diffusion coefficient smooths gradients. The sine-Gordon frequency continues to run according to
\begin{equation}
 \frac{\dd\beta}{\dd\ell}=-\frac{\beta^5}{8\pi}
 \int_0^{2\pi}\dd\theta\,[v'(\theta,\ell)]^2.
 \label{eq:beta-v}
\end{equation}
The relation to the reaction--diffusion variable is
\begin{equation}
 v=\frac{1}{2\pi}\frac{W'}{W},
 \label{eq:vW}
\end{equation}
so that
\begin{equation}
 W'=2\pi Wv.
 \label{eq:Wprime}
\end{equation}
Since $W$ is strictly positive, $u$ and $W$ have exactly the same extrema. The zeros of $v$, and especially its sign-changing zeros, therefore describe the creation or disappearance of maxima and minima of the running potential.

The essential property of Eq.~\eqref{eq:v} is its scalar parabolic structure~\cite{Henry1981}. Along any given smooth trajectory, its nonlinear term can be regarded as a drift with coefficient
\begin{equation}
 b(\theta,\ell)=\beta^2(\ell)v(\theta,\ell),
 \label{eq:drift}
\end{equation}
so that the same equation reads
\begin{equation}
 \partial_\ell v=D(\ell)v''+b(\theta,\ell)v'+2v.
 \label{eq:vparabolic}
\end{equation}
More explicitly, consider a compact RG interval on which the solution is smooth and bounded, $\beta(\ell)$ remains strictly positive, and the initial profile has finitely many zeros of $v$. Continuity then gives a positive lower bound on $D(\ell)=\beta^2(\ell)/(4\pi)$, while the drift $b=\beta^2v$ is bounded. These are the uniform-parabolicity and regularity conditions under which the classical one-dimensional zero-number result applies~\cite{Angenent1988}. The number of sign changes of $v$ cannot increase as $\ell$ grows. At a multiple zero this number can drop, reflecting a merger of neighboring extrema. Related coalescence events have been studied in viscous conservation laws~\cite{LinPelinovskyDeRijk2024}. Away from such events, simple zeros are transported smoothly and the extrema persist. In terms of the running potential this gives the geometric inequality
\begin{equation}
 N_{\mathrm{ext}}(\ell_2)\leq N_{\mathrm{ext}}(\ell_1),
 \qquad \ell_2\geq\ell_1,
 \label{eq:extrema-order}
\end{equation}
where $N_{\mathrm{ext}}$ counts sign-changing zeros of $u'$ (equivalently, nondegenerate extrema at generic RG scales). At a merger, it is preferable to use the sign-change count rather than to assign a number of extrema to a degenerate critical point. The simultaneous conditions $u'=u''=0$ are necessary for a merger, but they are not by themselves sufficient to establish a change in the number of extrema. A generic fold has $u'''\neq0$ at the event; a reflection-symmetric merger can instead exhibit a higher-order degeneracy, as discussed below.

The non-increase of extrema combines naturally with the preserved periodicity sectors obtained from the $W$ equation. A nonconstant potential with period $2\pi/p$ has at least one maximum and one minimum in each of its $p$ equivalent cells. Consequently, it must have at least $2p$ sign-changing extrema. If it starts with exactly that number, the symmetry imposes the lower bound $2p$ at later scales, while Eq.~\eqref{eq:extrema-order} imposes the same upper bound. The number of extrema is therefore fixed at $2p$ throughout any smooth, nonconstant finite-scale evolution. Additional pairs of extrema in the same symmetry sector are allowed to annihilate. In this precise sense, discrete translations protect the minimal geometry, although they do not freeze the detailed shape of the potential.

A complementary integral estimate illustrates how the same periodicity sectors enter the evolution of the amplitude of gradients. We introduce
\begin{equation}
 E(\ell)=\frac12\int_0^{2\pi}\dd\theta\,[v(\theta,\ell)]^2.
 \label{eq:E}
\end{equation}
Multiplication of Eq.~\eqref{eq:v} by $v$ and integration around the circle eliminates the advective contribution by periodicity, yielding
\begin{equation}
 \frac{\dd E}{\dd\ell}=4E-D\int_0^{2\pi}(v')^2\dd\theta.
 \label{eq:Eflow}
\end{equation}
When $v$ has period $2\pi/p$ and zero mean, the periodic Poincar\'e inequality implies
\begin{equation}
 \int_0^{2\pi}\dd\theta\,(v')^2\geq 2p^2E.
 \label{eq:poincare}
\end{equation}
Hence
\begin{equation}
 \frac{\dd E}{\dd\ell}\leq
 \left(4-\frac{p^2\beta^2}{2\pi}\right)E.
 \label{eq:bound}
\end{equation}
The coefficient changes sign at the $p$th Gaussian marginality threshold. This is an instantaneous estimate: because $\beta(\ell)$ itself evolves, it does not assert monotonic decay of $E$ along every RG trajectory. In contrast, the non-increase of the number of sign changes is a global statement for the parabolic projected flow and does not require a fixed $\beta$.

The two representations therefore address complementary aspects of coarse-graining. The reaction--diffusion equation makes the evolution of the full positive profile, its scaling dimensions, and its discrete symmetries transparent. The Burgers equation describes the zeros that encode the topology of its extrema. Taken together, they distinguish the possible generation of additional Fourier harmonics from the forbidden creation of new extrema within this approximation.

\section{Illustrations and discussion}
We consider two examples that illustrate how discrete translation symmetry
constrains the number of extrema along the projected Wilsonian flow.
In the first, a weak perturbation breaks the $\pi$-translation symmetry of
the initial potential. In the second, this symmetry is preserved throughout
the flow, but the initial potential has more extrema than the symmetry requires.

\subsection{Weakly broken translation symmetry}
Consider a potential proportional to $\cos(2\theta)$, with fundamental
period $\pi$. A small first harmonic breaks the symmetry under
$\theta\mapsto\theta+\pi$, giving the initial potential
\begin{equation}
 u(\theta,0)=2y_2\cos(2\theta)+2\varepsilon\cos\theta.
 \label{eq:double}
\end{equation}
When $\varepsilon=0$, the flow remains in $\Sp_2$, and its minimal four-extremum structure is protected by the preserved discrete translation symmetry. For $\varepsilon\neq0$, the exact $\pi$-translation symmetry is lost, although four extrema may initially remain. This setting therefore illustrates the zero-number constraint: once the symmetry protection is removed, the number of extrema may decrease, but it still cannot increase.

\begin{figure}[tbp]
 \centering
 \includegraphics[width=.85\linewidth]{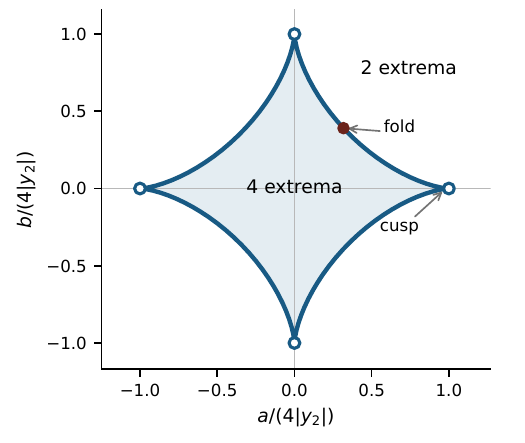}
 \caption{Astroid-like boundary for the two-harmonic family~\eqref{eq:ab}. Inside the curve are four extrema per $2\pi$ period; outside are two. Tips on the coordinate axes are reflection-symmetric cusp events (open circle denotes the tip reached for $b=0$, $a>0$); smooth boundary segments correspond to generic folds (filled marker). The diagram classifies potential shapes, not full RG trajectories.}
 \label{fig:astroid}
\end{figure}

To classify the initial shapes, the two-harmonic family can be extended from Eq.~\eqref{eq:double} to
\begin{equation}
 u(\theta)=2y_2\cos(2\theta)+2a\cos\theta+2b\sin\theta,
 \label{eq:ab}
\end{equation}
which allows one to classify the number of extrema in the $(a,b)$ plane. The boundary between the two-extremum and four-extremum regions is the astroid-like curve
\begin{equation}
 \left(\frac{|a|}{4|y_2|}\right)^{2/3}+\left(\frac{|b|}{4|y_2|}\right)^{2/3}=1.
 \label{eq:astroid}
\end{equation}
Inside this curve the potential has four extrema per $2\pi$ period, whereas outside it has only two. Figure~\ref{fig:astroid} shows this geometric classification of the two-harmonic family.  Along $b=0$, the criterion for four extrema reduces to $|a|<4|y_2|$; crossing either of the two tips on this axis is reflection symmetric. A generic crossing of a smooth segment of the astroid instead corresponds to an ordinary fold annihilation.

Along the full projected RG flow, however, additional harmonics are generated. The geometric merger is then characterized by the simultaneous conditions
\begin{equation}
 u'(\theta_*,\ell_*)=0,
 \label{eq:merger1}
\end{equation}
where the candidate extremum persists, together with
\begin{equation}
 u''(\theta_*,\ell_*)=0,
 \label{eq:merger2}
\end{equation}
which identifies a degenerate critical point. Whether the number of extrema actually changes is determined by the local structure and the direction of the flow. In the generic absence of reflection symmetry, $u'''(\theta_*,\ell_*)\neq0$ describes a fold: a neighboring maximum--minimum pair annihilates. In contrast, the initial condition~\eqref{eq:double} preserves reflection about $\theta=\pi$ even though its $\pi$-translation symmetry is broken. This enforces $u'(\pi,\ell)=u'''(\pi,\ell)=0$ at every scale. At the observed event, $u''(\pi,\ell_{\mathrm{geom}})=0$ while $u^{(4)}(\pi,\ell_{\mathrm{geom}})\neq0$, so that
\begin{equation}
 u'(\pi+s,\ell_{\mathrm{geom}})
 =\frac{u^{(4)}(\pi,\ell_{\mathrm{geom}})}{6}s^3+O(s^5).
 \label{eq:cubic-merger}
\end{equation}
The zero of $u'$ is therefore cubic: two off-axis extrema meet a symmetry-pinned extremum and only the latter survives. In the two-harmonic picture, this corresponds to crossing a cusp along $b=0$ rather than a generic fold.

%\sout{Figure~\ref{fig:profiles} shows the same $4\to2$ event in the full projected flow with harmonics generated dynamically. The initial conditions are $y_2(0)=0.01$, $\varepsilon=10^{-8}$, and $\beta^2(0)=7.6290209582$. With 14 positive Fourier harmonics, the curvature criterion $u''(\pi,\ell_{\mathrm{geom}})=0$ locates the event at $\ell_{\mathrm{geom}}\simeq9.0471$. The three profiles correspond to $\Delta\ell\equiv\ell-\ell_{\mathrm{geom}}=-0.55,0,+0.55$. The Fourier--Galerkin equations were integrated with an adaptive DOP853 solver ($\mathrm{rtol}=2\times10^{-10}$, $\mathrm{atol}=10^{-12}$, maximum RG step $0.12$), and the symmetric merger was identified by the zero of $u''(\pi,\ell)$. No new extrema appear as the off-axis pair is lost. The astroid classifies the two-harmonic shapes, whereas this flow illustrates the corresponding allowed geometric simplification in a dynamically generated multi-harmonic potential. The merger scale depends on the RG truncation and the initial parameters.}

\begin{figure}[tbp]
 \centering
 \includegraphics[width=.85\linewidth]{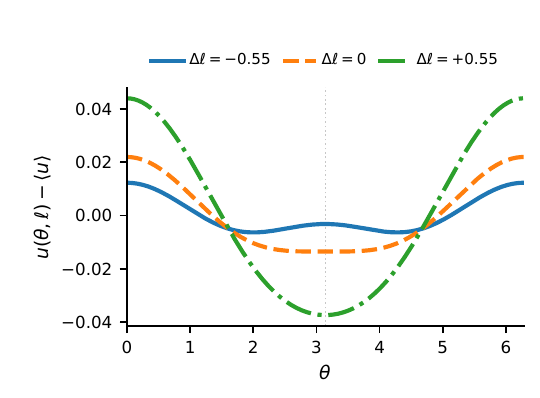}
 \caption{
%	\sout{Four-to-two extrema along the projected Wilsonian flow, using 14 positive Fourier harmonics, $y_2(0)=0.01$, $\varepsilon=10^{-8}$, and $\beta^2(0)=7.6290209582$. The mean of $u$ is subtracted. Profiles at $\ell-\ell_{\mathrm{geom}}=-0.55,0,+0.55$ bracket the reflection-symmetric cubic merger at $\theta=\pi$; $\ell_{\mathrm{geom}}\simeq9.0471$ is found from $u''(\pi,\ell)=0$. Distinct line styles aid grayscale reading. The accompanying numerical script reproduces the curves.}
 Four-to-two reduction of extrema along the projected Wilsonian flow. The initial potential is given by Eq.\,\eqref{eq:double}, with \(y_2(0)=0.01\), \(\varepsilon=10^{-8}\), and \(\beta^2(0)=7.6290209582\). The profiles are shown at three RG scales relative to the merger scale, \(\Delta\ell\equiv\ell-\ell_{\mathrm{geom}}=-0.55,\,0,\,+0.55\), where \(\ell_{\mathrm{geom}}\simeq 9.0471\) is determined by the condition \(u''(\pi,\ell_{\mathrm{geom}})=0\). 
 At the merger, two off-axis extrema meet the symmetry-pinned extremum at \(\theta=\pi\), producing a cubic zero of \(u'\). The resulting potential has two extrema per period. The flow is computed using the Fourier-Galerkin integration method with 14 positive Fourier harmonics; the mean of \(u\) is subtracted from each profile for clarity.}
 \label{fig:profiles}
\end{figure}

Figure~\ref{fig:profiles} illustrates the $4\to2$ reduction of extrema obtained by numerically integrating the functional flow equation \eqref{eq:u}. We solve the equation using a pseudospectral method, representing the periodic potential as a truncated Fourier series and evaluating the nonlinear terms in Fourier space. The resulting Fourier–Galerkin system is integrated in the RG scale using an adaptive DOP853 solver, retaining 14 positive Fourier harmonics. The initial conditions are ($y_2(0)=0.01$), ($\varepsilon=10^{-8}$), and ($\beta^2(0)=7.6290209582$).

The merger occurs at ($\ell_{\mathrm{geom}}\simeq 9.0471$), identified by the condition ($u''(\pi,\ell_{\mathrm{geom}})=0$). The three profiles shown in Fig. 2 correspond to RG scales before, at, and after this event, with ($\Delta\ell\equiv\ell-\ell_{\mathrm{geom}}=-0.55,,0,,+0.55$), respectively. As predicted by the zero-number property, no new extrema are generated during the evolution; instead, the two off-axis extrema merge with the symmetry-pinned extremum, leaving only two extrema per period.

\subsection{Preserved translation symmetry}
Consider a non-sinusoidal potential with fundamental period $\pi$.
To examine how the relative amplitudes of two harmonics affect the
number of extrema within each period, we choose the initial profile
\begin{equation}
 u(\theta,0)=g\ln\!\left[1+a\cos(2\theta)+b\cos(4\theta)\right],
 \label{eq:log-profile}
\end{equation}
with the argument of the logarithm strictly positive.

Although the logarithm generally contains infinitely many Fourier harmonics, its extrema can be counted analytically because
\begin{equation}
 u'(\theta,0)=
 -\frac{2g\sin(2\theta)[a+4b\cos(2\theta)]}
 {1+a\cos(2\theta)+b\cos(4\theta)}.
 \label{eq:log-derivative}
\end{equation}
For nonzero $g$ and $b$, and away from the degeneracy $|a|=4|b|$, the condition $\sin(2\theta)=0$ supplies four stationary points over the angular circle. If $|a|<4|b|$, four additional ones occur at $\cos(2\theta)=-a/(4b)$, giving eight nondegenerate extrema in all. When $|a|>4|b|$, only the four symmetry-pinned extrema remain.

\begin{figure}[t]
 \centering
 \includegraphics[width=.98\linewidth]{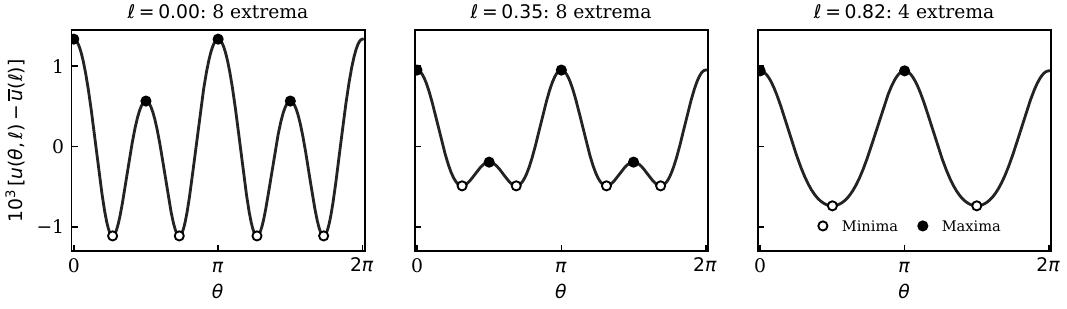}
 \caption{Symmetry preserving loss of excess extrema for the non-sinusoidal initial potential~\eqref{eq:log-profile}, with $g=0.004$, $a=0.12$, $b=0.25$, and $\beta^2(0)=1.2\pi$. The three profiles at $\ell=0$, $0.35$, and $0.82$ show the reduction from eight to four extrema across $\ell_{\mathrm{geom}}\simeq0.61882$. Filled and open circles denote maxima and minima, respectively; the mean of $u$ is subtracted. The $\pi$-translation symmetry remains exact throughout the projected flow. Each profile is computed with 24 positive even Fourier harmonics; $\theta=0$ and $2\pi$ represent the same point on the circle.}
 \label{fig:log-profile}
\end{figure}

We evolve Eq.~\eqref{eq:log-profile} under the same projected functional flow, starting from $g=0.004$, $a=0.12$, $b=0.25$, and $\beta^2(0)=1.2\pi$. The logarithm is everywhere real and the initial profile has eight extrema. The exact translation symmetry $u(\theta+\pi,\ell)=u(\theta,\ell)$ is maintained throughout the evolution. Figure~\ref{fig:log-profile} illustrates the ensuing loss of four excess extrema, leaving precisely the four protected by this symmetry. With 24 positive even Fourier harmonics, the merger takes place at $\ell_{\mathrm{geom}}\simeq0.61882$, determined from $u''(\pi/2,\ell_{\mathrm{geom}})=0$. The two symmetry related coalescences at $\theta=\pi/2$ and $3\pi/2$ are cubic, rather than generic folds. Varying the number of retained positive even harmonics between 8 and 30 changes the merger scale by less than $10^{-8}$. The initial $n=2$ and $n=4$ scaling exponents are $0.8$ and $-2.8$, respectively, providing a simple weak-coupling explanation of the geometric simplification. The numerical trajectory follows the full projected potential equation, including its generated harmonics and running $\beta$.

The two trajectories illustrate complementary consequences of the geometric constraints. Breaking the $\pi$-translation symmetry permits a reduction from four extrema to two, whereas preserving it does not prevent eight extrema from reducing to the symmetry-protected minimum of four. The invariant periodicity sector thus protects a lower bound, not the entire initial landscape.

The global results are appropriately qualified. An equation for a complete periodic function is not an exact RG equation: the derivation retains terms through second order in the interaction and projects away field-dependent kinetic operators generated at that order. Consequently, the non-increase of the sign change count is established \emph{within this parabolic projected flow}; it need not survive arbitrary improvements of the truncation. Appendix~\ref{app:polchinski} shows that nonconstant kinetic harmonics are linearly irrelevant near the Gaussian fixed line, but are nevertheless generated at quadratic order. A comparison with the coupled potential kinetic factor equations finds the same $4\to2$ merger in the weak coupling examples considered, with shifted merger scales. This numerical observation does not establish the same sign change property for the coupled flow. The shape of individual trajectories and the location of merger scales are nonuniversal. These limitations should be distinguished from the symmetry of the underlying microscopic model, which is preserved by any symmetry-respecting RG.

\section{Conclusions}

We have investigated how the shape of a general smooth periodic interaction evolves under Wilsonian coarse graining in two dimensions. Within the functional flow considered here, the reaction diffusion and Burgers representations reveal two complementary properties: discrete translation symmetries define invariant sectors, and the number of extrema of the running potential can only decrease or remain constant. The latter follows from the zero-number property of scalar parabolic equations. An interesting consequence is that the nonlinear generation of Fourier harmonics can accompany a progressive simplification of the potential landscape. The geometric evolution therefore provides information that complements the conventional description in terms of harmonic scaling.

The interplay between symmetry and geometric simplification is illustrated by two distinct examples. A weak breaking of the \(\pi\)-translation symmetry allows a potential with four extrema to evolve into one with only two. In contrast, a smooth non-sinusoidal potential that retains this symmetry evolves from eight extrema to four, reaching the minimum number protected by the invariant periodicity sector. Together, these examples show how coarse-graining reorganizes the competing minima and barriers of a periodic interaction, while its exact symmetries constrain the resulting landscape.

These results suggest several directions for further investigation. A natural extension is to explore how the geometric constraints evolve when field dependent kinetic terms are retained in the renormalization group flow, building on the coupled-flow analysis presented in the Appendix. It would also be interesting to examine their implications for multifrequency sine-Gordon theories, commensurate structures, and substrate pinning models, where competing minima play an essential role. More generally, the evolution of extrema offers a complementary perspective on how complex periodic interactions are reorganized across scales.

\section*{Acknowledgments}
We acknowledge partial financial support from the Brazilian funding agencies \emph{Fundação Carlos Chagas Filho de Amparo à Pesquisa do Estado do Rio de Janeiro} (FAPERJ), \emph{Conselho Nacional de Desenvolvimento Científico e Tecnológico} (CNPq), and \emph{Coordenação de Aperfeiçoamento de Pessoal de Nível Superior} (CAPES, Finance Code 001). We also acknowledge support from the INCT project \emph{Advanced Quantum Materials}, supported by CNPq (Grant No.~408766/2024-7), FAPESP (Grant No.~2025/27091-3), and CAPES, as well as from the CAPES--COFECUB program.

D.G.B. thanks the \emph{Laboratoire de Physique Théorique et Hautes Énergies} (LPTHE) at Sorbonne Université in Paris for its warm hospitality during his stay as a visiting professor, when this work was completed.N.S. acknowledge funding by the Spanish
Ministerio de Ciencia, Innovación y Universidades (MICIU/AEI/10.13039/501100011033) through the Maria
de Maeztu project CEX2021-001164-M.

\appendix
\addtocontents{toc}{\protect\fixappendix}

\section{Polchinski RG, wave-function renormalization,
and the field-independent kinetic projection}

\label{app:polchinski}

\subsection{Matching the functional equations}

To compare our blocking calculation with a different differential RG formulation, we consider the derivative-expanded, dimensionless action
\begin{equation}
 S_k[\varphi]=\int\dd^2x\left[\frac12 Z_k(\varphi)(\nabla\varphi)^2+V_k(\varphi)\right].
 \label{eq:app-action}
\end{equation}
The potential $V_k$ and wave-function factor $Z_k$ may both depend periodically on the field. We use the notation of N\'{a}ndori \emph{et al.}~\cite{Nandori2002} for the regulator-dependent coefficients,
\begin{equation}
 I_1=\int\frac{\dd^2q}{(2\pi)^2}\,q^2 K'(q^2),
 \label{eq:app-I1}
\end{equation}
\begin{equation}
 I_2=\int\frac{\dd^2q}{(2\pi)^2}\,K'(q^2),
 \label{eq:app-I2}
\end{equation}
where the dimensionless integration momentum is expressed in units of $k$ and $K(s)$ is the smooth regulator. Primes on $K$ denote differentiation with respect to its argument, whereas subscripts $\varphi$ below denote field derivatives.

At order $O(\partial_x^2)$ in the derivative expansion, the Polchinski equations can be written as
\begin{align}
 (2+k\partial_k)V_k={}&-K'_0(V_{k,\varphi})^2
 +(Z_k-K_0^{-1})I_1+I_2 V_{k,\varphi\varphi},
 \label{eq:app-pol-V}
\end{align}
\begin{align}
 k\partial_k Z_k={}&-4K'_0(Z_k-K_0^{-1})V_{k,\varphi\varphi}
 -2K'_0 V_{k,\varphi}Z_{k,\varphi}\notag\\
 &-2K''_0(V_{k,\varphi\varphi})^2+I_2 Z_{k,\varphi\varphi}.
 \label{eq:app-pol-Z}
\end{align}
The equations correspond to Eq.~(15) of Ref.~\cite{Nandori2002}, with $\ell=\ln(\Lambda/k)$ increasing toward the infrared. For our Gaussian choice $K(s)=e^{-s}$,
\begin{equation}
 K_0=1,\qquad K'_0=-1,\qquad K''_0=1,
 \label{eq:app-K}
\end{equation}
and direct radial integration gives
\begin{equation}
 I_1=I_2=-\alpha,\qquad \alpha\equiv\frac{1}{4\pi}.
 \label{eq:app-I}
\end{equation}
Consequently, the coupled equations become
\begin{equation}
 \partial_\ell V=2V+\alpha V_{\varphi\varphi}-(V_\varphi)^2
 +\alpha(Z-1),
 \label{eq:app-Vell}
\end{equation}
\begin{align}
 \partial_\ell Z={}&\alpha Z_{\varphi\varphi}
 +2(V_{\varphi\varphi})^2\notag\\
 &-2V_\varphi Z_\varphi-4(Z-1)V_{\varphi\varphi}.
 \label{eq:app-Zell}
\end{align}
These equations retain the field dependence of the kinetic factor; our closed two-variable flow is obtained by an additional projection described below.

\subsection{Projection onto a constant wave-function factor}

We split the kinetic factor into an angular average and a zero-mean periodic component,
\begin{equation}
 Z(\theta,\ell)=Z_0(\ell)+\widetilde Z(\theta,\ell),
 \qquad \langle\widetilde Z\rangle_\theta=0,
 \label{eq:app-Zsplit}
\end{equation}
where $\langle f\rangle_\theta=(2\pi)^{-1}\int_0^{2\pi}f(\theta)\,\dd\theta$. Within the field-independent kinetic projection, the oscillatory component is set to zero after each infinitesimal blocking step; the average $Z_0$ is retained and subsequently absorbed by a linear field rescaling. Before this rescaling we use
\begin{equation}
 V(\varphi,\ell)=-\frac12 u(\beta\varphi,\ell),
 \qquad\theta=\beta\varphi,
 \label{eq:app-map}
\end{equation}
with $\beta$ fixed within the blocking step and $Z=1$. The potential equation~\eqref{eq:app-Vell} then reads
\begin{equation}
 \partial_\ell u=2u+\frac{\beta^2}{4\pi}u''
 +\frac{\beta^2}{2}(u')^2,
 \label{eq:app-u}
\end{equation}
in exact agreement with Eq.~\eqref{eq:u} at the order and projection specified in the main text.

For the kinetic factor, Eq.~\eqref{eq:app-Zell} generates at $Z=1$
\begin{equation}
 \left.\partial_\ell Z(\theta)\right|_{Z=1}
 =2(V_{\varphi\varphi})^2
 =\frac{\beta^4}{2}[u''(\theta)]^2.
 \label{eq:app-Zsource}
\end{equation}
Its constant part is therefore
\begin{equation}
 \partial_\ell Z_0=\frac{\beta^4}{2}
 \big\langle [u''(\theta)]^2\big\rangle_\theta.
 \label{eq:app-Z0}
\end{equation}
Rescaling the field to restore a unit kinetic coefficient gives $\beta\mapsto\beta/\sqrt{1+\delta Z_0}$, so that
\begin{equation}
 \frac{\dd\beta}{\dd\ell}=-\frac{\beta}{2}\,
 \frac{\dd Z_0}{\dd\ell}
 =-\frac{\beta^5}{4}\big\langle(u'')^2\big\rangle_\theta.
 \label{eq:app-beta}
\end{equation}
This is Eq.~\eqref{eq:beta}. Thus, for a Gaussian cutoff and the same constant-kinetic projection, the cumulant and Polchinski calculations yield identical coefficients.

\subsection{Irrelevance of nonconstant kinetic harmonics}

The reason for using a constant $Z_0$ is best understood by linearizing near the Gaussian fixed line $V=0$, $Z=1$. Expanding the zero-mean part as
\begin{equation}
 \widetilde Z(\theta,\ell)=\sum_{n\neq0}z_n(\ell)e^{in\theta},
 \label{eq:app-FourierZ}
\end{equation}
Eq.~\eqref{eq:app-Zell} yields, at fixed Gaussian $\beta$,
\begin{equation}
 \frac{\dd z_n}{\dd\ell}
 =-\frac{n^2\beta^2}{4\pi}z_n+O(Vz,V^2).
 \label{eq:app-zlinear}
\end{equation}
Hence each nonconstant kinetic harmonic is irrelevant in the Gaussian linearization, in agreement with Ref.~\cite{Nandori2002}. For example, a single-harmonic potential $u=2y\cos(p\theta)$ generates
\begin{equation}
 \partial_\ell Z\big|_{Z=1}
 =\beta^4p^4y^2[1+\cos(2p\theta)].
 \label{eq:app-singleZ}
\end{equation}
At the $p$th KT marginality threshold~\eqref{eq:Wthresholds}, the generated $2p$ kinetic harmonic has the negative linear eigenvalue $\lambda_{Z,2p}=-8$.

This provides a useful \emph{near-Gaussian, infrared} motivation for projecting onto $Z_0$. It is not, however, an order-by-order perturbative identity: Eq.~\eqref{eq:app-singleZ} shows that the constant and oscillatory kinetic terms are both generated at order $y^2$. In the unprojected equation~\eqref{eq:app-Vell}, $\widetilde Z$ in turn feeds back into $V$ at the same formal order. Thus the flow with a field-independent kinetic factor is a specified truncation with linearly irrelevant omitted operators.

\subsection{Coupled-flow check and limitations of the zero-number statement}

For a numerical sensitivity test we use a scheme in which the mean kinetic factor is restored to unity at each step. We write $Z(\theta,\ell)=1+z(\theta,\ell)$ with $\langle z\rangle_\theta=0$. Equations~\eqref{eq:app-Vell} and~\eqref{eq:app-Zell} then give the coupled system
\begin{equation}
 \partial_\ell u=2u+Du''+\frac{\beta^2}{2}(u')^2-2\alpha z,
 \label{eq:app-extended-u}
\end{equation}
\begin{align}
 \partial_\ell z={}&Dz''+\frac{\beta^4}{2}(u'')^2
 +\beta^2u'z'\notag\\
 &+2\beta^2zu''-\eta(1+z),
 \label{eq:app-extended-z}
\end{align}
where
\begin{equation}
 \eta=\frac{\beta^4}{2}\langle(u'')^2\rangle_\theta
 +\beta^2\langle zu''\rangle_\theta,
 \label{eq:app-eta}
\end{equation}
\begin{equation}
 \frac{\dd\beta}{\dd\ell}=-\frac{\eta}{2}\beta.
 \label{eq:app-extended-beta}
\end{equation}
Here $D=\alpha\beta^2$, and the subtraction proportional to $\eta$ enforces $\langle z\rangle_\theta=0$. Setting $z=0$ \emph{as a projection after each step} recovers the flows used in the main text. The zero-mean component is not an invariant subspace of the unprojected system.

We integrated both systems for the initial data
\begin{equation}
 u(\theta,0)=0.02\cos(2\theta)+2\varepsilon\cos(\theta-\varphi),
 \quad z(\theta,0)=0,
 \label{eq:app-numericalIC}
\end{equation}
with identical $\beta^2(0)=7.6290209582$ and a Fourier--Galerkin discretization. Table~\ref{tab:app-numerics} lists the merger scales for four illustrative runs. Spectral convergence checks used 12, 18, 28, and 40 positive harmonics. Across these four resolutions, the projected merger times in the reflection-symmetric tests with $\varepsilon=10^{-4}$ and $10^{-6}$ vary by less than $10^{-6}$, while the corresponding extended-flow values vary by less than $1.5\times10^{-6}$. These bounds apply to the tabulated convergence tests. In all the cases tabulated, both systems exhibit a $4\to2$ reduction, but retaining $z$ shifts the location of the event.
\begin{table}[tbp]
\centering
\caption{Numerical merger scales for equal initial $\beta^2$, with and without the periodic kinetic component. The last column gives $\max_\theta|z|$ at the event in the extended flow. Here $\ell_Z$ is the merger scale with the periodic kinetic component retained. All entries correspond to $4\to2$ extrema.}
\label{tab:app-numerics}
\footnotesize
\setlength{\tabcolsep}{3.4pt}
\begin{tabular}{@{}cccc@{}}
\toprule
$(\varepsilon,\varphi)$ & $\ell_{\rm proj}$ & $\ell_{Z}$ & $\max|z|$ \\
\midrule
$(10^{-4},0)$ & 3.3806 & 3.5934 & 0.0172 \\
$(10^{-6},0)$ & 6.1770 & 6.5364 & 0.0110 \\
$(10^{-8},0)$ & 9.0471 & 9.5606 & 0.0093 \\
$(10^{-4},0.37)$ & 3.1420 & 3.2175 & 0.0111 \\
\bottomrule
\end{tabular}
\end{table}

Because the near-KT trajectory is sensitive to the initial Gaussian stiffness, we also tuned the two models separately in the symmetric sector $\varepsilon=0$. Our \emph{finite-scale} approximate separatrix prescription imposes $x+Y=0$ at $\ell=20$, with $x=2-\beta^2/\pi$ and $Y=8\sqrt{2}\pi|y_2|$ (12 retained positive harmonics in the tuning). It gives $\beta^2(0)\simeq7.62864$ for the projected flow and $7.71459$ for the extended flow. With $\varepsilon=10^{-8}$, the resulting merger scales are $9.0477$ and $9.3605$, respectively. This finite-scale criterion is not an exact determination of the KT separatrix; the numerical merger scales are scheme dependent and these examples do not prove a global zero-number property for the extended system.

The discrete translation sectors remain invariant in Eqs.~\eqref{eq:app-extended-u}--\eqref{eq:app-extended-beta}, because both $u$ and $z$ inherit any common initial periodicity. In contrast, differentiating Eq.~\eqref{eq:app-extended-u} introduces the extra term $-2\alpha z'$ into the equation for $u'$. The scalar homogeneous parabolic zero-number argument used in the main text no longer applies directly. Consequently, the numerical persistence of the $4\to2$ example supports its robustness in the weak-coupling cases tested, but the \emph{no-creation property remains established only for the projected functional flow}.

%\bibliographystyle{apsrev4-2}
%\bibliography{references}


\end{document}